\documentclass[letterpaper,english,aps, nofootinbib, pr, twocolumn, lengthcheck, superscriptaddress]{revtex4-2}

\usepackage[T1]{fontenc}
\usepackage[colorlinks=true,linktocpage=true,linkcolor=blue,citecolor=blue]{hyperref}
\usepackage{color,txfonts}
\usepackage{verbatim}
\usepackage{bm}
\usepackage{amssymb}
\makeatletter
\let\iint\@undefined
\let\iiint\@undefined
\let\iiiint\@undefined
\let\idotsint\@undefined
\makeatother
\usepackage{amsmath}
\usepackage{graphicx}
\usepackage{slashed}
\usepackage{wrapfig}
\usepackage[caption=false]{subfig}
\usepackage{verbatim}
\usepackage{enumitem}

\begin{document}

\title{Post-Merger Gravitational-Wave Uncertainties of Binary Neutron Stars under Multi-Messenger EOS Constraints}

\author{Yong-Jia Huang} 
\email{huangyj@pmo.ac.cn}
\affiliation{Key Laboratory of Dark Matter and Space Astronomy, Purple Mountain Observatory, Chinese Academy of Science, Nanjing, 210023, China.}
\affiliation{RIKEN Center for Interdisciplinary Theoretical and Mathematical Sciences (iTHEMS), RIKEN, Wako 351-0198, Japan}

\author{Luca Baiotti}
\affiliation{International College and Graduate School of Science, The University of Osaka, 1-2 Machikaneyama-cho, Toyonaka, Osaka 560-0043, Japan}

\begin{abstract}
The high-frequency gravitational waves emitted by a binary neutron star merger remnant carry information on matter at densities and temperatures beyond those reached in isolated neutron stars. We quantify how tightly current multi-messenger constraints already determine the dominant post-merger frequency $f_{2,\rm mean}$. Adopting a set of cold equations of state (EOSs) constrained jointly by gravitational-wave tidal deformability, NICER mass--radius measurements, massive-pulsar masses, chiral effective field theory at low density, and perturbative QCD at asymptotically high density, for each binary mass we select the softest and stiffest models of the multi-messenger posterior and follow their coalescence with fully general-relativistic hydrodynamics simulations. Together with a broad set of EOSs drawn from the literature ($82$ models in total), these simulations show that, once the binary mass and a single measure of the stellar compactness ($\Lambda$ or $R$) are held fixed, the residual spread of $f_{2,\rm mean}$ is only $\sim 100\,{\rm Hz}$, a factor of several below the $\gtrsim 500\,{\rm Hz}$ range spanned by an EOSs set including those already disfavored by the data. This tight calibration of the cold-matter prediction implies that future high-frequency detections departing from it would point directly to additional physics, such as a hadron--quark transition occurring at finite temperature. We further confirm the quasi-universal relation $(f_1+f_3)/2 \approx f_{2,\rm mean}$ to within $\sim 116\,{\rm Hz}$, which provides a model-independent estimate of $f_{2,\rm mean}$ from the secondary spectral peaks.
\end{abstract}

\maketitle

\section{Introduction}
\label{sec:intro}

The coalescence of binary neutron stars (BNS) offers a unique laboratory for probing the properties of matter at supranuclear densities. Gravitational-wave (GW) observations of the inspiral phase (GW170817~\cite{Abbott2017} and GW190425~\cite{Abbott2020}), together with electromagnetic constraints from the NICER mission~\cite{Miller2019, Riley2019, Miller2021, Riley2021}, have significantly narrowed the allowed parameter space of the equation of state (EOS) of cold, dense matter. These multi-messenger data have proven instrumental in constraining the EOS up to several times nuclear saturation density. However, the EOS at the highest densities ({\it i.e.}, in the innermost core) has only a weak impact on global observables such as the mass--radius relation and tidal deformability. As a result, different high-density EOSs (including those featuring a hadron--quark phase transition) can produce neutron stars with very similar macroscopic properties, limiting what can be inferred about the deep-core EOS from observations of stable neutron stars alone~\citep{Drischler-etal_2021PrPNP}. The post-merger phase, instead, has the potential to probe dense matter at more extreme temperatures and densities that exceed those realized in stable neutron stars. This could be achieved by studying the remnant's dynamics via its high-frequency GW emission.

Numerical-relativity simulations indicate that, if the remnant is a temporarily or long-lived stable neutron star, the post-merger spectrum is dominated by several peaks, including a dominant peak, with frequency named $f_2$, which correlates robustly with the tidal properties of the progenitor stars~\cite{Bauswein2012, Bernuzzi2015, Rezzolla2016, Breschi2019}. Moreover, comparing the observed post-merger GW frequency with the value predicted from the inspiral signal using a given EOS can distinguish between different EOS models more effectively than constraints from stable neutron stars alone~\cite{Huang:2022mqp,Hensh:2024onv}.

The $f_2$ spectral feature is a primary target for next-generation GW observatories~\citep{Prakash2024PhRvD109}, such as the Einstein Telescope~\cite{Punturo2010} and Cosmic Explorer~\cite{Reitze2019}, which aim to overcome the limited high-frequency sensitivity of current detectors. A key question for the upcoming era of precision GW astronomy is to quantify to what extent our current knowledge of the EOS, derived from inspiral and pulsar data, already constrains these future post-merger observables.

Previous studies of BNS mergers have often relied on a limited set of phenomenological EOS models~\cite{Vretinaris-2020PhRvD.101, Prakash:2021wpz, Haque:2022dsc, Huang:2022mqp, Fujimoto:2022xhv, Most:2022wgo, Blacker:2023afl, Harada:2023eyg, Hensh:2024onv, Han:2025pho, Tsokaros-MHD-2025PhRvL.134, Bamber-MHD-2025PhRvD.111, Tsokaros-2025arXiv251017511R, Magnall-2026MNRAS.tmp.1130M} and, in some cases, have focused on numerical and physical uncertainties associated with the simulations ({\it e.g.}, code systematics, grid resolution, magnetic fields, bulk or shear viscosity, and finite-temperature effects). However, adopting a small set of cold EOS models does not necessarily capture the latest multi-messenger constraints. To robustly forecast the observability of post-merger signals, it is essential to employ an inference framework that minimizes phenomenological bias while respecting fundamental physical principles. Because of the steep computational cost of numerical-relativity simulations, a direct Bayesian exploration of the EOS at the level routinely carried out for stable neutron stars (see, e.g., \cite{Han:2022rug,Gorda:2022jvk,Huang:2026jgl,Grundler:2026zus,Tang:2026yta,Gorda:2025aiu} is currently infeasible for BNS mergers, since a single merger evolution ($\sim 10\,$days of wall-clock time on a supercomputer) is roughly eight orders of magnitude more expensive than a single EOS likelihood evaluation in the stable-star problem ($\sim 10\,$ms).

In this work, we investigate uncertainties in post-merger GW emission by using a flexible, non-parametric Bayesian inference approach for the EOS, without assuming a specific microphysical model. The EOSs are constructed by exploring the full functional space allowed by causality, thermodynamic stability, and perturbative QCD (pQCD) constraints at asymptotically high densities. We condition this broad ensemble on the latest multi-messenger observations, combining GW tidal measurements with updated mass--radius constraints from massive pulsars and NICER.

By mapping these rigorously constrained EOS posteriors to the post-merger regime via established quasi-universal relations, we determine the uncertainty in the peak frequency across a range of binary masses, up to the {\it no-$f_2$} threshold~\cite{Hensh:2024onv}. In practice, for each given binary mass, we identify bounding "soft" and "stiff" models by selecting the EOSs corresponding to the lower and upper edges of the credible region of the remnant's central density; the resulting bounding EOSs can therefore vary with the total mass.

We find that, once the binary mass and a pre-merger compactness proxy (the tidal deformability $\Lambda$ or the radius $R$) are fixed, the dominant post-merger frequency $f_{2,\rm mean}$ is predicted to within a weighted root-mean-square uncertainty of $\sigma_{f_2}\simeq100\,{\rm Hz}$ across the observationally allowed EOS ensemble, a factor of several tighter than the $\gtrsim500\,{\rm Hz}$ spread spanned by EOSs that are no longer favored by current data. We also find that this residual uncertainty is comparable to the frequency shift ($\sim100$--$120\,{\rm Hz}$) induced by varying the thermal index $\Gamma_{\rm th}$ between $1.6$ and $2.0$. Consequently, a future post-merger detection at $\lesssim100\,{\rm Hz}$ precision, which is within reach of the Einstein Telescope and Cosmic Explorer for nearby events, combined with an inspiral measurement of $(M,\Lambda)$, would directly probe the finite-temperature behavior of supranuclear matter, a regime inaccessible to inspiral and pulsar observations of cold, stable neutron stars.

\section{Method}
\label{sec:method}
We generate quasi-equilibrium irrotational binary initial data using the multi-domain pseudo-spectral code \texttt{Lorene}~\citep{Gourgoulhon2001} with an initial separation of $45\,\text{km}$. The general-relativistic hydrodynamics is evolved with \texttt{WhiskyTHC}~\citep{10.1093/mnrasl/slt137, Radice:2013xpa} within the \texttt{Einstein Toolkit}~\citep{EinsteinToolkit:2023_05}. The numerical scheme for the evolution of the hydrodynamics equations uses a finite-volume approach with fifth-order monotonicity-preserving reconstruction and the HLLE Riemann solver. Spacetime is evolved in the Z4c formulation with "1+log" slicing and Gamma-driver shift conditions. Time integration employs the method of lines with a third-order strong-stability-preserving Runge--Kutta scheme and a Courant factor of $0.075$. Adaptive mesh refinement is provided by \texttt{Carpet}~\citep{ErikSchnetter_2004} with seven levels, yielding a finest grid spacing of $\approx 231\,\text{m}$; the outer boundary is placed at $1477\,\text{km}$.

\begin{figure}[htbp]
\centering
\vspace{-0.3cm}
\includegraphics[width=0.5\textwidth]{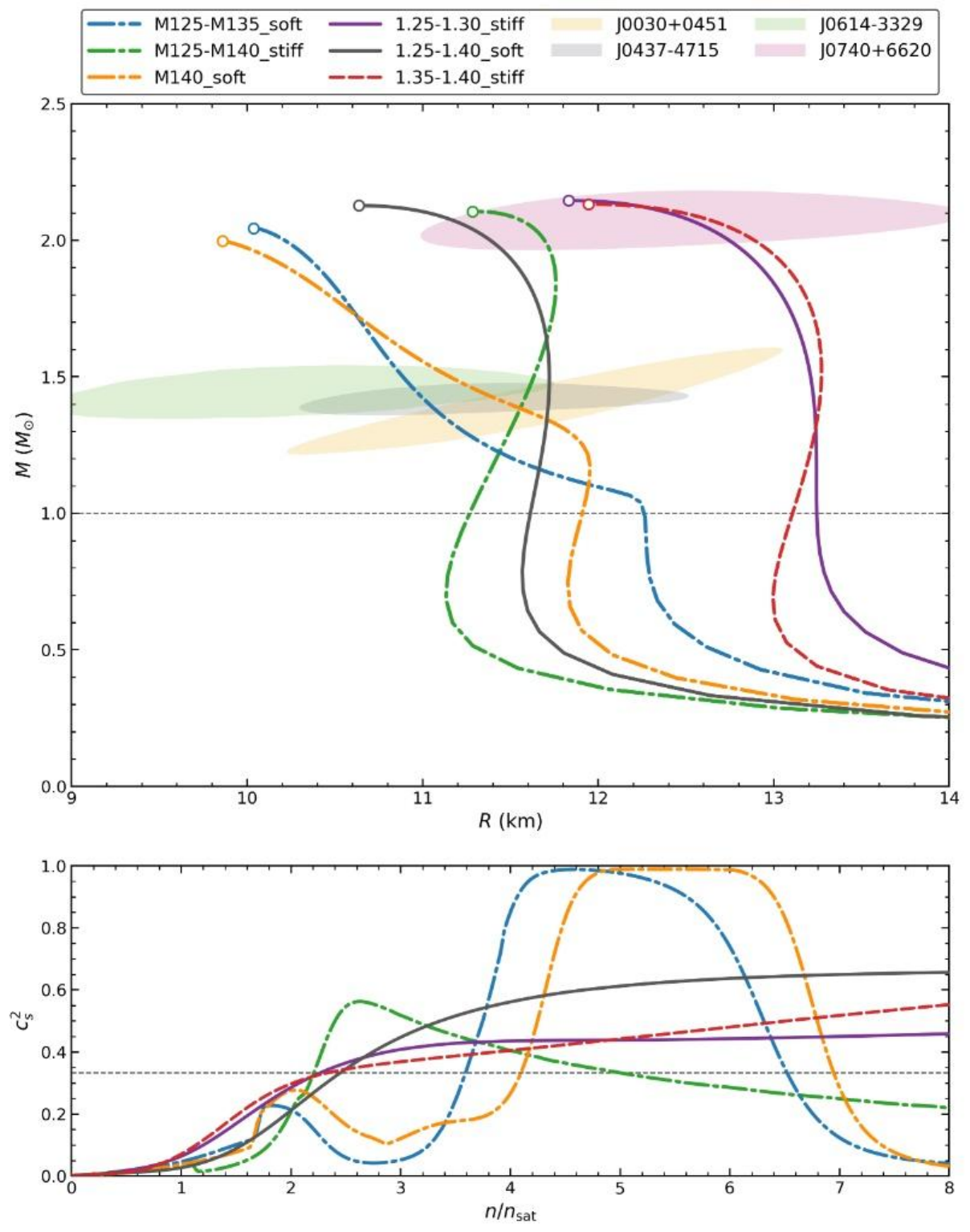}
\vspace{-0.3cm}
\caption{Mass-radius relations and squared sound speed ($c_s^2$) for statistical EOS models.
Models from the FFNN construction are selected if their radius at a given mass falls within the 90\% credible region, while models from~\citep{Huang:2026jgl} (which incorporate additional information from PSR~J0437$-$4715 and PSR~J0614$-$3329) are selected using the 68\% highest posterior density interval of the joint posterior.}
\label{fig:selected_models}
\end{figure}

We consider two classes of non-parametric statistical equations of state (EOSs). The first class is constructed by first generating EOSs with a feed-forward neural network (FFNN) method~\citep{Han_2021ApJ...919...11H, Han:2022rug}, then retaining only those that fall within the 90\% credible region of the multi-messenger parameter constraints, and, finally,  selecting, for each binary mass, the models with the smallest and largest radius as our {\it soft} and {\it stiff} (extremal) models. The second class adopts the approach of~\citep{Huang:2026jgl}, which builds EOSs by connecting continuously a chiral-effective-field-theory EOS at low density to a perturbative QCD one at asymptotically high density. This analysis incorporates the recent NICER measurements for PSR~J0437$-$4715 and PSR~J0614$-$3329, and we use the 68\% highest posterior density interval to select a set of EOSs exhibiting a wider range of high-density behavior.
For each mass between $1.25\,M_{\odot}$ and $1.40\,M_{\odot}$, we then identify the stiffest and softest models based on their central density. The selected models span the observational uncertainties in neutron-star properties. The two stiff models and the soft model from~\citep{Han:2022rug} are referred to as 1.25-1.30\_stiff, 1.35-1.40\_stiff, and 1.25-1.40\_soft, while the two soft models and the stiff model from~\citep{Huang:2026jgl} are labeled M125-M135\_soft, M140\_soft, and M125-M140\_stiff. The mass--radius relations and squared sound speeds $c_s^2$ are shown in Fig.~\ref{fig:selected_models}.

Since shock heating brings the merger remnant to temperatures of tens of MeV, thermal effects dominate the post-merger dynamics. Although microphysical three-parameter tables $P(n, T, Y_{\rm e})$ offer a more realistic description of such hot environments, constructing them for our non-parametric EOSs would require unconstrained phenomenological assumptions, as multi-messenger observations constrain only dense matter in cold NSs. To avoid introducing model-dependent information into our statistical framework, we instead adopt the commonly used {\it ideal-gas} approximation~\citep{ideal-fluid-2010PhRvD82}. The total pressure is given by $P = P_{\rm cold} + P_{\rm th}$, where $P_{\rm cold}$ is taken from the EOS table and the thermal pressure is $P_{\rm th} = (\Gamma_{\rm th} - 1) \cdot \epsilon_{\rm th}$. The thermal internal energy is obtained from $\epsilon_{\rm th} = \epsilon - \epsilon_{\rm cold}$, with $\epsilon$ determined by the hydrodynamics equations. In this framework, thermal effects are governed by the adiabatic index $\Gamma_{\rm th}$, which typically ranges from $1.5$ to $2.0$ in realistic cases~\citep{ideal-fluid-2010PhRvD82}. In this work, we set $\Gamma_{\rm th} = 2$ for the models from~\citep{Han:2022rug} and consider $\Gamma_{\rm th} = 1.6$ and $2$ for the models from~\citep{Huang:2026jgl}. This control-variable approach effectively brackets the leading-order thermal uncertainties while providing a conservative, model-independent bound on the post-merger frequency variations.

\section{Result}
\label{sec:result}
The dynamical fate of the post-merger remnant ({\it i.e.}, a nearly prompt collapse or longer survival) determines whether a detectable $f_2$ signal is produced, and is itself shaped by the structural properties of the EOS. Before describing the simulation outcomes, we therefore highlight a key feature of the Huang~et~al.\ 2026~\citep{Huang:2026jgl} (hereafter "updated") EOS ensemble on which our analysis is based. The pQCD constraints, which require the sound speed to approach the conformal limit $c_s^2 \to 1/3$ at asymptotically high baryon densities~\citep{Kurkela-2010PhRvD..81j5021K, Gorda-2021PhRvL.127p2003G}, result in a systematic anti-correlation between the EOS stiffness at low and high densities (Fig.~\ref{fig:selected_models}): A model that is \emph{soft} at sub-nuclear to nuclear densities (small radius $R$, small tidal deformability $\Lambda$, i.e.\ our {\it soft updated model}) must compensate by being \emph{stiffer} at supranuclear densities in order to support observed massive pulsars ($M_{\rm max} \gtrsim 2\,M_\odot$); conversely, a model that is stiff at nuclear densities (large $R$, large $\Lambda$) is forced by the pQCD asymptote to soften at the highest densities. As a result, the currently allowed EOS space, constrained simultaneously by multi-messenger observations at low-to-intermediate densities and by pQCD at high densities, cannot remain uniformly soft or stiff across the full density range probed in a BNS merger.

\begin{figure}[htbp]
\centering
\vspace{-0.3cm}
\includegraphics[width=0.95\columnwidth]{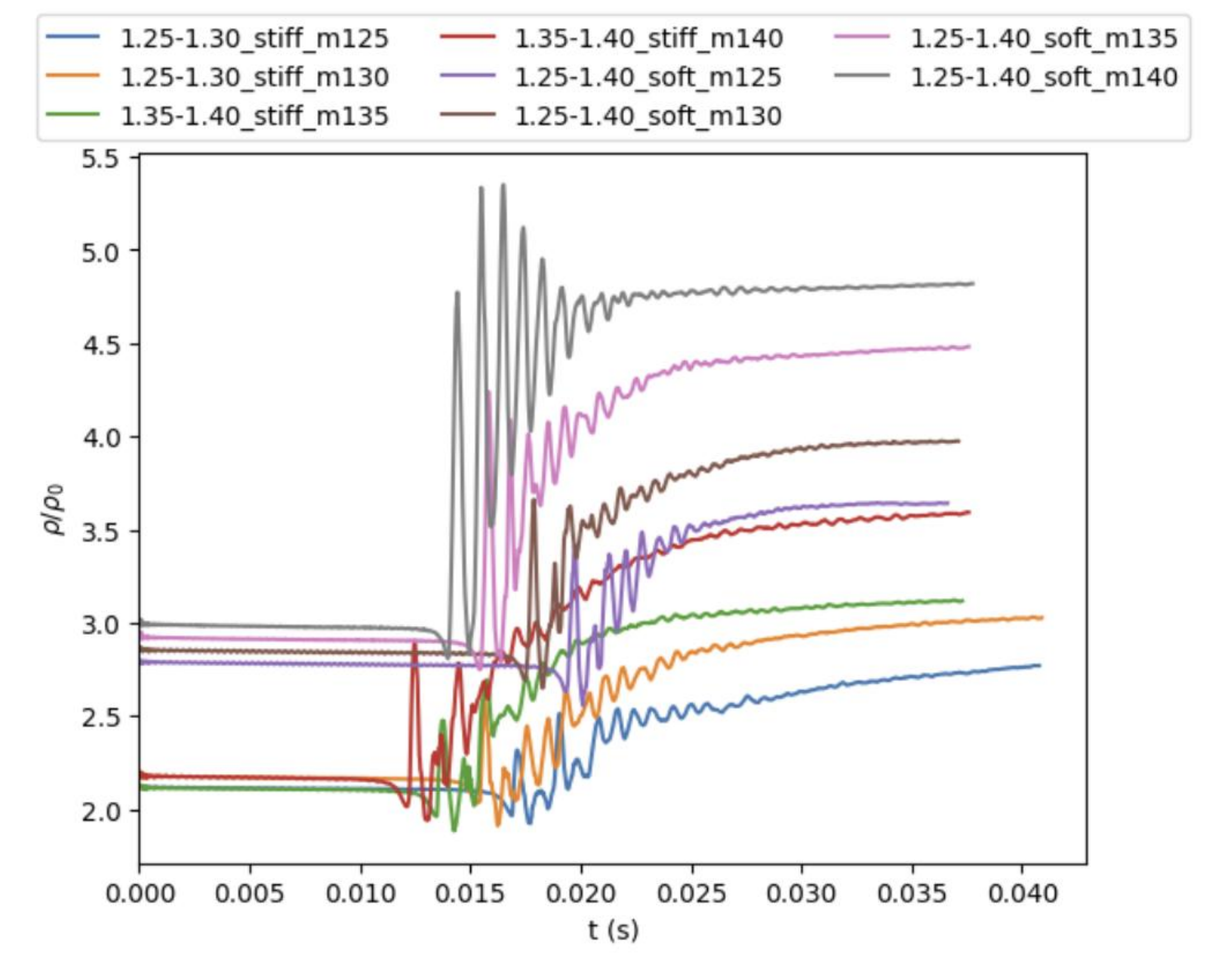}
\vspace{-0.3cm}
\caption{Time evolution of the maximum rest-mass density $\rho_{\rm max}$, in units of the nuclear saturation density $\rho_0$, for the eight Han23~\citep{Han:2022rug} configurations evolved with $\Gamma_{\rm th}=2.0$. Following merger at $t\approx12$--$18\,{\rm ms}$, $\rho_{\rm max}$ relaxes into stable quasi-periodic oscillations about a quasi-equilibrium value for all four mass configurations ($M_{\rm NS}=1.25$--$1.40\,M_\odot$), signalling the formation of a long-lived massive neutron-star remnant in every case.}
\label{fig:rho_max_old}
\end{figure}

Figure~\ref{fig:rho_max_old} shows the time evolution of the maximum rest-mass density $\rho_{\rm max}$ for all Han~et~al.\ 2023 \citep{Han:2022rug} (hereafter "Han23")
models. Since the FFNN produces simple $c_s^2$ structures in the Han23 model construction and we selected the models based on their radii, the cases for both small and large radii are described by a monotonic sound speed. Combined with their relatively large radii ($R \approx 11.7$--$13.2\,{\rm km}$), these models provide strong support against collapse, and $\rho_{\rm max}$ oscillates stably around a quasi-equilibrium value in all four mass configurations ($M_{\rm NS} = 1.25$--$1.40\,M_\odot$, where $M_{\rm NS}$ is the baryon mass of one ov the stars in the equal-mass binary), indicating the formation of a long-lived remnant.

\begin{figure}[htbp]
\centering
\includegraphics[width=0.95\columnwidth]{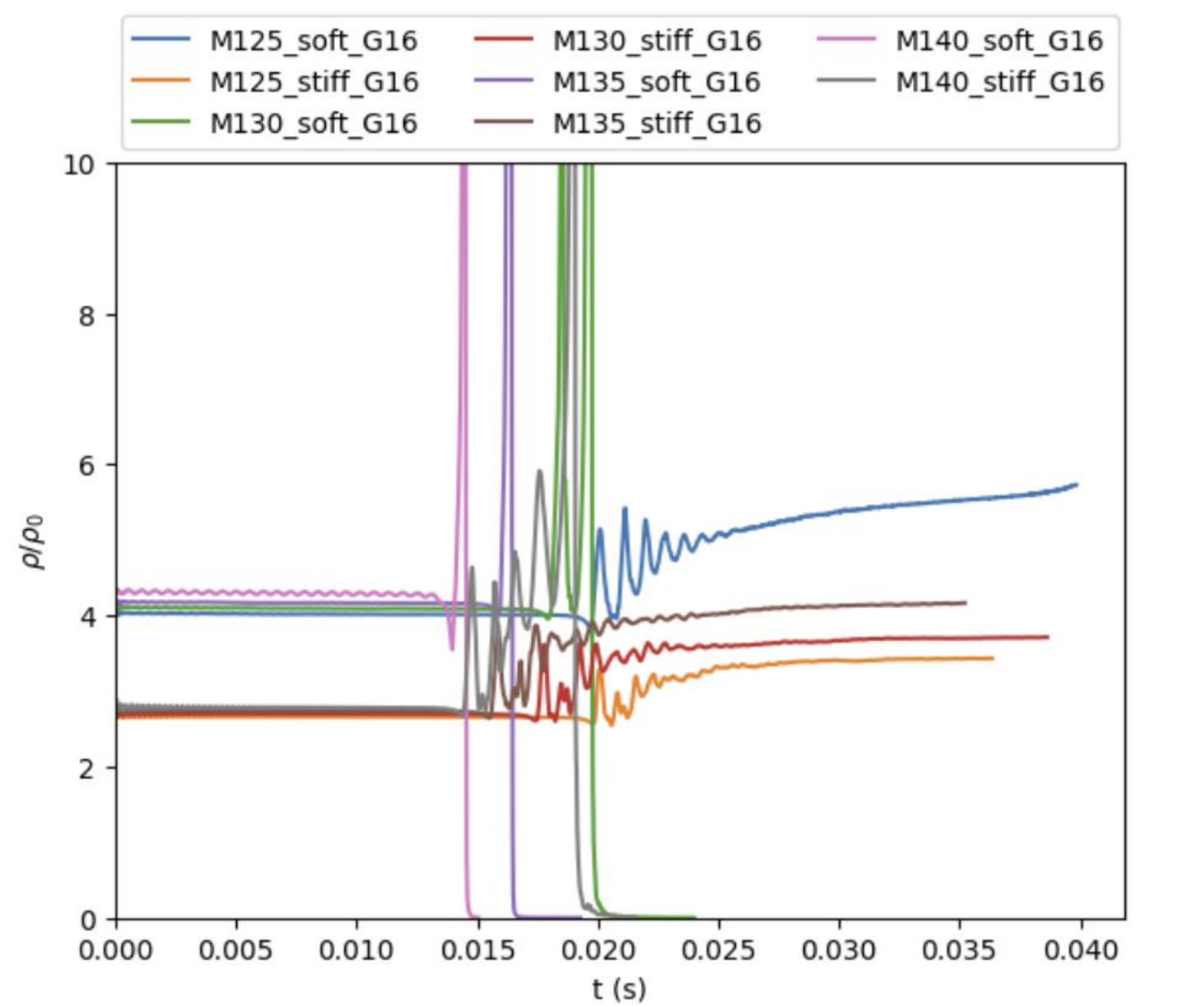}
\vspace{0.2cm} 
\includegraphics[width=0.95\columnwidth]{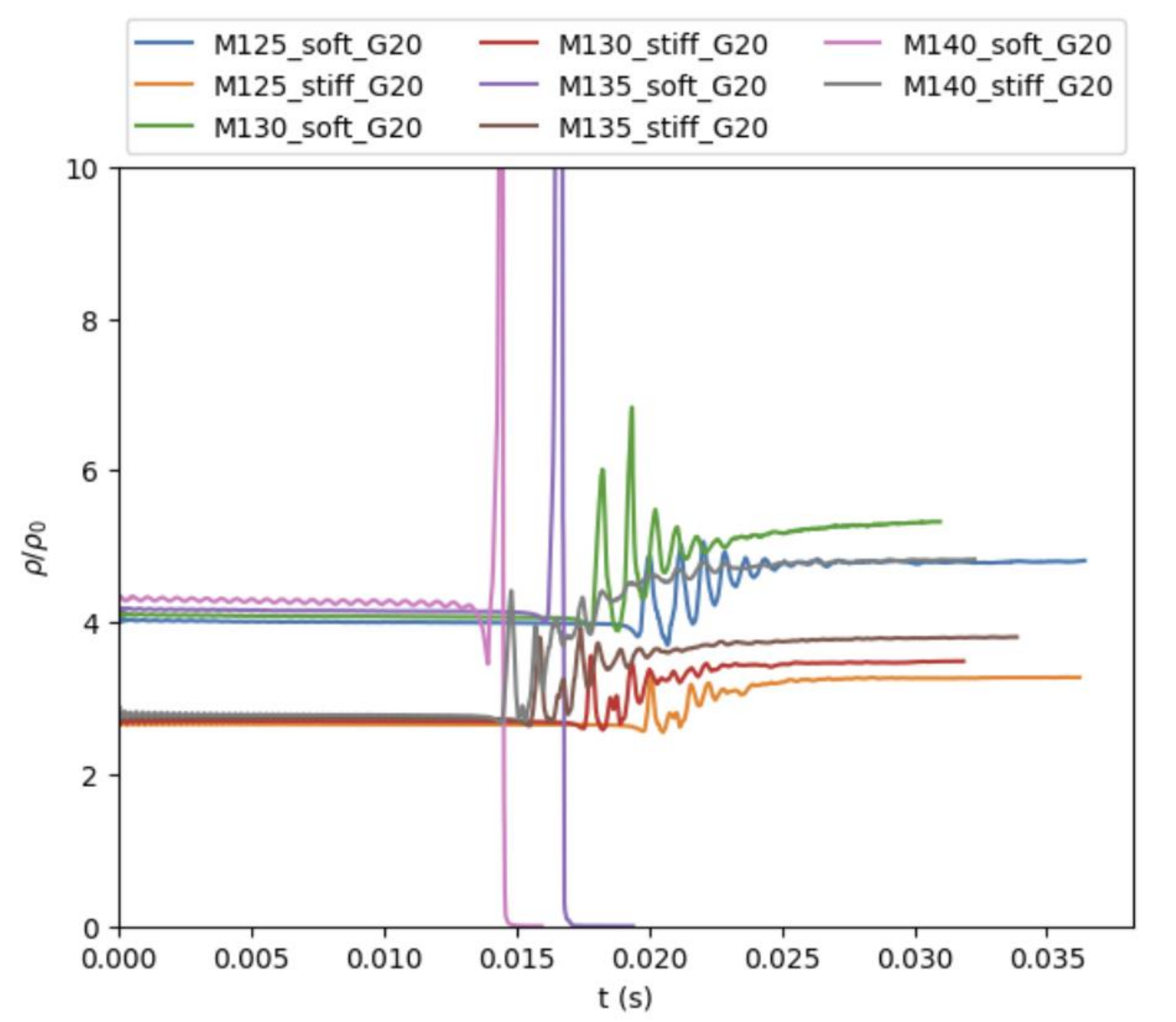}
\caption{As in Fig.~\ref{fig:rho_max_old}, but for the updated configurations~\citep{Huang:2026jgl} evolved with thermal index $\Gamma_{\rm th}=1.6$ (top) and $\Gamma_{\rm th}=2.0$ (bottom). The stiff updated EOS produces a stable remnant at every mass considered here and for both thermal prescriptions, whereas the soft updated EOS causes collapse to a black hole (this is signalled by the abrupt drop of $\rho_{\rm max}$ to zero following a transient spike) at a certain threshold binary mass, which is smaller for smaller $\Gamma_{\rm th}$. In particular, the $M_{\rm NS}=1.30\,M_\odot$ soft model survives only for $\Gamma_{\rm th}=2.0$, illustrating how the weaker thermal pressure support at lower $\Gamma_{\rm th}$ reduces the effective maximum mass of the hot remnant.}
\label{fig:rho_max_new}
\end{figure}

Figure~\ref{fig:rho_max_new} presents the corresponding results for the updated models with $\Gamma_{\rm th}=1.6$ and $2.0$. For the stiff updated models ($R \approx 11.46$--$11.57\,{\rm km}$, $\Lambda \approx 317$--$578$), stable remnants are produced at all masses and for both $\Gamma_{\rm th}$ values. The soft updated models ($R \approx 11.23$--$11.36\,{\rm km}$, $\Lambda \approx 296$--$409$) are more vulnerable to collapse. At $M_{\rm NS} = 1.25\,M_\odot$ both $\Gamma_{\rm th}$ values yield long-lived remnants; at $M_{\rm NS} = 1.30\,M_\odot$ only $\Gamma_{\rm th}=2.0$ gives a detectable $f_2$ signal while $\Gamma_{\rm th}=1.6$ undergoes rapid collapse; and at $M_{\rm NS} \geq 1.35\,M_\odot$ both $\Gamma_{\rm th}$ values lead to prompt or near-prompt collapse with no recoverable $f_2$ peak. A smaller $\Gamma_{\rm th}$ provides less post-merger thermal pressure support, lowering the effective maximum mass of the hot remnant; for the already-compact soft EOS, this shift is sufficient to cross the stability boundary at progressively lower binary masses, whereas the stiff EOS models, being farther from their stability limit, remain unaffected by the $\Gamma_{\rm th}$ variation within the range studied.

The observational consequence is a reduction of the viable EOS space. As shown in Fig.~\ref{fig:rho_max_new}, for $M_{\rm total} \gtrsim 2.70\,M_\odot$, any merger that produces a detectable $f_2$ signal must be consistent with the stiffer updated EOS branch, thereby reducing the viable EOS space.

For all surviving remnants, the post-merger GW spectrum exhibits up to three characteristic peaks: the dominant quadrupolar (bar-mode) frequency $f_2$~\citep{Sterg-2011MNRAS.418}, and two minor peaks $f_1$ and $f_3$~\citep{Takami-Rezzolla-Baiotti-2014PhRvL} (see also~\citep{Sterg-2011MNRAS.418}). The radial oscillation frequency $f_0$, which is extracted from the oscillations of $\rho_{\rm max}$ and does not itself appear as a peak in the GW spectrum, provides the frequency scale for the approximate relations $f_1 \approx f_2 - f_0$ and $f_3 \approx f_2 + f_0$~\citep{Sterg-2011MNRAS.418,Takami-Rezzolla-Baiotti-2014PhRvL, Takami-Rezzolla-Baiotti-PhRvD91, Bauswein-Stergioulas-2015PhRvD91}.
Although these secondary modes are widely attributed to a non-linear coupling between the bar mode and the radial oscillation, the precise mechanism has not yet been definitively established. The frequencies extracted from all models are listed in Table~\ref{tab:allmodels}. In some cases, the $f_3$ data is left blank because the $f_3$ mode cannot be distinguished from the background.

\begin{table*}
  \centering
\caption{Properties of the simulated BNS merger systems. The listed quantities are the gravitational mass ($M_\textrm{NS}$) of one isolated star, the baryon mass ($M_\textrm{b,NS}$), the total angular momentum $J$, the stellar radius $R$, 
the dimensionless tidal deformability $\Lambda$,
the contact frequency ($f_\textrm{cont} = C^{3/2}/2\pi M$ where $C=M/R$), the radial oscillation frequency $f_0$, and the post-merger frequencies $f_1$, $f_{2,\textrm{mean}}$, and $f_3$ with their 68\% confidence intervals. All simulations use our medium resolution ($\Delta x \approx 231\,\text{m}$).}
\vspace{0.35cm}
  \resizebox{\textwidth}{!}{\begin{tabular}{lcccccccccc}
    \hline\hline \\
    Model & $M_\textrm{NS}$[$M_\odot$] & $M_\textrm{b,NS}$[$M_\odot$] & $J$[$M_\odot^2$] & $R$[km] & $\Lambda$ & $f_\textrm{cont}$[Hz] & $f_0$[Hz] & $f_1$[Hz] & $f_{2,\textrm{mean}}$[Hz] & $f_3$[Hz] \\ \\
    \hline \\[-0.5em]
    1.25-1.30-stiff-m125 & 1.25 & 1.35 & 6.40 & 13.23 & 1226 & 1499 & 1152 & $1830^{+10}_{-10}$ & $2458^{+10}_{-6}$ & $2910^{+9}_{-9}$ \\[0.5em]
    1.25-1.30-stiff-m130 & 1.30 & 1.41 & 6.83 & 13.23 & 980  & 1565 & 1124 & $2071^{+109}_{-88}$ & $2664^{+55}_{-40}$ & -- \\[0.5em]
    1.35-1.40-stiff-m135 & 1.35 & 1.47 & 7.27 & 13.24 & 863  & 1617 & 1072 & $1910^{+10}_{-9}$ & $2530^{+7}_{-7}$ & $2998^{+7}_{-7}$ \\[0.5em]
    1.35-1.40-stiff-m140 & 1.40 & 1.53 & 7.73 & 13.25 & 701  & 1504 & 1090 & $1926^{+20}_{-18}$ & $2709^{+8}_{-7}$ & $3680^{+10}_{-10}$ \\[0.5em]
    1.25-1.40-soft-m125 & 1.25 & 1.37 & 6.40 & 11.68 & 611  & 1570  & 1310 & $2215^{+36}_{-28}$ & $3215^{+20}_{-23}$ & $4353^{+13}_{-13}$ \\[0.5em]
    1.25-1.40-soft-m130 & 1.30 & 1.43 & 6.83 & 11.69 & 490  & 1623 & 1266 & $2294^{+16}_{-16}$ & $3219^{+10}_{-9}$ & $4362^{+9}_{-9}$ \\[0.5em]
    1.25-1.40-soft-m135 & 1.35 & 1.49 & 7.26 & 11.70 & 393  & 1653 & 1197 & $2402^{+25}_{-24}$ & $3331^{+13}_{-13}$ & $4438^{+11}_{-11}$ \\[0.5em]
    1.25-1.40-soft-m140 & 1.40 & 1.55 & 7.72 & 11.71 & 318  & 1605 & 1085 & $2432^{+21}_{-16}$ & $3290^{+28}_{-17}$ & $4332^{+12}_{-9}$ \\[0.5em]
    M125-softG16  & 1.25 & 1.38 & 6.47 & 11.36 & 409 & 1693 & 1129 & $2921^{+16}_{-16}$ & $3936^{+8}_{-8}$ & $5012^{+5}_{-6}$ \\[0.5em]
    M125-softG20  &  &  &  &  &  &  & 988 & $2737^{+15}_{-14}$ & $3649^{+8}_{-8}$ & $4744^{+11}_{-10}$ \\[0.5em]
    M125-stiffG16 & 1.25 & 1.38 & 6.47 & 11.46 & 578 & 1671 & 1347 & $2467^{+21}_{-19}$ & $3270^{+16}_{-15}$ & -- \\[0.5em]
    M125-stiffG20 &  &  &  &  &  &  & 1351 & $2340^{+12}_{-10}$ & $3276^{+14}_{-13}$ & $4466^{+11}_{-11}$ \\[0.5em]
    M130-softG16  & 1.30 & 1.45 & 7.00 & 11.23 & 296 & 1756 & 908 & -- & -- & -- \\[0.5em]
    M130-softG20  & & &  & & &  & 1033 & $3044^{+14}_{-15}$ & $3920^{+13}_{-13}$ & $4927^{+7}_{-7}$ \\[0.5em]
    M130-stiffG16 & 1.30 & 1.44 & 6.90 & 11.50 & 470 & 1696 & 1404 & $2572^{+35}_{-36}$ & $3373^{+16}_{-17}$ & $4512^{+18}_{-17}$ \\[0.5em]
    M130-stiffG20 &  &  &  &  &  &  & 1380 & $2395^{+13}_{-14}$ & $3253^{+7}_{-18}$ & $4485^{+47}_{-46}$ \\[0.5em]
    M135-softG16  & 1.35 & 1.51 & 7.45 & 11.04 & 165 & 1837 & -- & -- & -- & -- \\[0.5em]
    M135-softG20  &  &  &  &  &  &  & -- & -- & -- & -- \\[0.5em]
    M135-stiffG16 & 1.35 & 1.50 & 7.34 & 11.53 & 383 & 1720 & 1305 & $2619^{+156}_{-115}$ & $3468^{+22}_{-20}$ & $4655^{+13}_{-14}$ \\[0.5em]
    M135-stiffG20 &  &  &  &  &  &  & 1348 & $2343^{+23}_{-23}$ & $3357^{+16}_{-16}$ & $4572^{+12}_{-12}$ \\[0.5em]
    M140-softG16  & 1.40 & 1.57 & 7.87 & 11.49 & 250 & 1762 & -- & -- & -- & -- \\[0.5em]
    M140-softG20  &  &  &  &  &  &  & -- & -- & -- & -- \\[0.5em]
    M140-stiffG16 & 1.40 & 1.56 & 7.78 & 11.57 & 317 & 1744 & -- & $2472^{+40}_{-34}$ & $3392^{+32}_{-26}$ & $4557^{+26}_{-25}$ \\[0.5em]
    M140-stiffG20 &  &  &  &  &  &  & 1115 & $2446^{+19}_{-16}$ & $3375^{+18}_{-17}$ & $4611^{+14}_{-15}$ \\[0.5em]
    \hline\hline
\end{tabular}}
  \label{tab:allmodels}
\end{table*}

At fixed M, the mean frequency of the dominant GW peak $f_{2,\rm mean}$ anti-correlates with the progenitor radius $R$ (and with the tidal deformability $\Lambda$), since more compact progenitors form more compact, more rapidly rotating remnants that radiate GW at higher frequencies. Among the Han23 models, the stiff--soft spread at fixed mass reaches $\sim 560$--$760\,{\rm Hz}$, driven by the large differences in $\Lambda$ ($\Lambda \approx 701$--$1226$ for stiff vs.\ $318$--$611$ for soft). The primary ($\Gamma_{\rm th}=2.0$) stiff updated models give $f_{2,\rm mean} = 3253$--$3375\,{\rm Hz}$ over $M_{\rm NS} = 1.30$--$1.40\,M_\odot$, while the soft updated models, where they survive, reach $3649\,{\rm Hz}$ (M125-softG20) and $3920\,{\rm Hz}$ (M130-softG20). Comparing the supplementary $\Gamma_{\rm th}=1.6$ runs to the primary $\Gamma_{\rm th}=2.0$ results for the stiff updated EOS reveals a systematic thermal shift, in which lower $\Gamma_{\rm th}$ yields a more compact remnant and higher $f_{2,\rm mean}$, with shifts of $6$, $120$, $111$, and $17\,{\rm Hz}$ at $M_{\rm NS} = 1.25$, $1.30$, $1.35$, $1.40\,M_\odot$, respectively. Therefore, the $f_{2,\rm mean}$ shift from thermal effect is also EOS-dependent.

These systematics reflect three physical mechanisms that together determine $f_{2,\rm mean}$ and motivate the structure of the quasi-universal fitting formula we now introduce. The first and most immediate mechanism is the role of binary mass; for a given EOS a heavier binary produces a denser remnant, so $f_{2,\rm mean}$ increases with $M$ at fixed EOS stiffness. The second mechanism operates through the pre-merger stellar compactness encoded in $R$ or $\Lambda \propto (R/M)^5$. A more compact star (smaller $R$, smaller $\Lambda$) has a higher contact frequency $f_{\rm cont} = C^{3/2}/(2\pi M)$ (where $C = M/R$; see Table~\ref{tab:allmodels})~\citep{Bauswein2012, Takami-Rezzolla-Baiotti-2014PhRvL, Rezzolla2016}, meaning that the two stars are orbiting faster when they merge. Because the initial rotation frequency of the remnant is set by the orbital angular momentum at contact, a compact pre-merger configuration seeds a faster-rotating remnant and hence a higher $f_2$. This second mechanism is accessible from inspiral measurements, as both the mass--radius relation (from NICER) and the tidal deformability (from the GW inspiral phase) probe the NS structure at intermediate densities $\rho_c \sim 2$--$5\,\rho_{\rm sat}$, characteristic of stable $1.2$--$2\,M_\odot$ stars. A qualitatively distinct, third mechanism enters at densities reached only transiently during and after merger, far above $2\,\rho_{\rm sat}$ and inaccessible to inspiral observations. The response of the high-density EOS at these extreme conditions, which is not encoded in $R$ or $\Lambda$, directly modulates how much the remnant further compresses post-merger and hence how fast it oscillates or rotates~\citep{Huang:2022mqp, Hensh:2024onv}. In the updated models, the pQCD-enforced anti-correlation described above gives this mechanism a specific character. The soft updated model (small $R$, soft at nuclear density) carries a \emph{stiffer} supranuclear EOS that stiffens the remnant beyond what the pre-merger compactness alone would predict, partially offsetting the $f_2$ enhancement expected from its smaller $R$; conversely, the stiff updated model (large $R$) has a softer high-density EOS, reducing the remnant compactness below the naive expectation from its larger radius. Because this cross-density trade-off cannot be captured by $\Lambda$ or $R$ alone, it contributes a systematic residual in any fit based solely on pre-merger observables.

Over the parameter range covered by the observationally allowed EOS ensemble, the dependence of $f_2$ on $M$, $R$, and $\ln\Lambda$ (all of which are accessible from isolated NS observations) is seen to be approximately linear. 
With the purpose of removing the leading dependence on mass and compactness so that the remaining dispersion can be interpreted physically, we adopt the following two-variable fitting formulas
\begin{equation}
    f_2^{\rm fit} = \alpha_0 + \alpha_M\,(M - 1.4\,M_\odot) + \alpha_X\,g(X),
    \label{eq:fit}
\end{equation}
one with $X = R$ with $g(R) = R - R_{1.4}$, and the other with $X = \Lambda$ with $g(\Lambda) = \ln(\Lambda/\Lambda_{1.4})$. By including $M$ (first mechanism) and $X$ (dominant part of the second mechanism) as predictors, the fit simultaneously captures the leading dependence of $f_2$ on the two dominant physical mechanisms. Such empirical relations, formulated via alternative combinations of progenitor masses and structural properties, have been widely explored in the literature~\cite{Bauswein2012, Bernuzzi2015, Rezzolla2016, Vretinaris-2020PhRvD.101}. The residual $f_{2,\rm mean}$ prediction uncertainty, defined as the weighted root-mean-square error (RMSE) $\sigma_{f_2} \equiv \bigl(\sum_i w_i\,e_i^2 / \sum_i w_i\bigr)^{1/2}$ where $e_i = f_{2,\rm mean}^{(i)} - f_2^{{\rm fit},(i)}$ is the deviation of the $i$-th simulation model from the fit, directly 
provides an upper bound on the contribution of the third mechanism, which encompasses supranuclear EOS stiffness together with any higher-order dependences not captured by the linear $M$--$X$ parameterization. The fit is performed by Gaussian-weighted least squares, where the weight of each simulation model is defined as:
\begin{equation}
    w = \exp(-z^2/2),
    \label{eq:weight}
\end{equation}
with the squared distance measure
\begin{equation}
    z^2 = \frac{1}{2}\left[ \left(\frac{R-R_{1.4}}{\sigma_R}\right)^2 + \left(\frac{\ln(\Lambda/\Lambda_{1.4})}{\sigma_\Lambda}\right)^2 \right],
    \label{eq:z2}
\end{equation}
so that $\sigma_{f_2}$ reflects the prediction uncertainty for an EOS drawn from the observationally allowed posterior, not the full spread over all conceivable EOSs.  

The full primary sample comprises 82 models from four families: 8 Han23 models and 6 updated models from our numerical-relativity simulations (all with $\Gamma_{\rm th}=2.0$), together with 43 hadronic (APR4, SLy, togashi, GNH3, H4, ALF2)\cite{Rezzolla2016} and 25 quark--hadron crossover (QHC19B/C/D, QHC21A/AT/D/DT)\cite{Huang:2022mqp,Hensh:2024onv} models compiled from the literature. The five $\Gamma_{\rm th}=1.6$ updated runs share the same cold EOS as their $\Gamma_{\rm th}=2.0$ counterparts and are therefore not counted as independent samples; they are retained only as supplementary points in the mass-corrected analysis below. The ensemble spans $R \approx 11.0$--$13.9\,{\rm km}$ and $f_{2,\rm mean} \approx 2000$--$3950\,{\rm Hz}$, providing a broad lever arm for the regression.

\begin{figure*}[htbp]
\centering
\vspace{-0.3cm}
\includegraphics[width=1\textwidth]{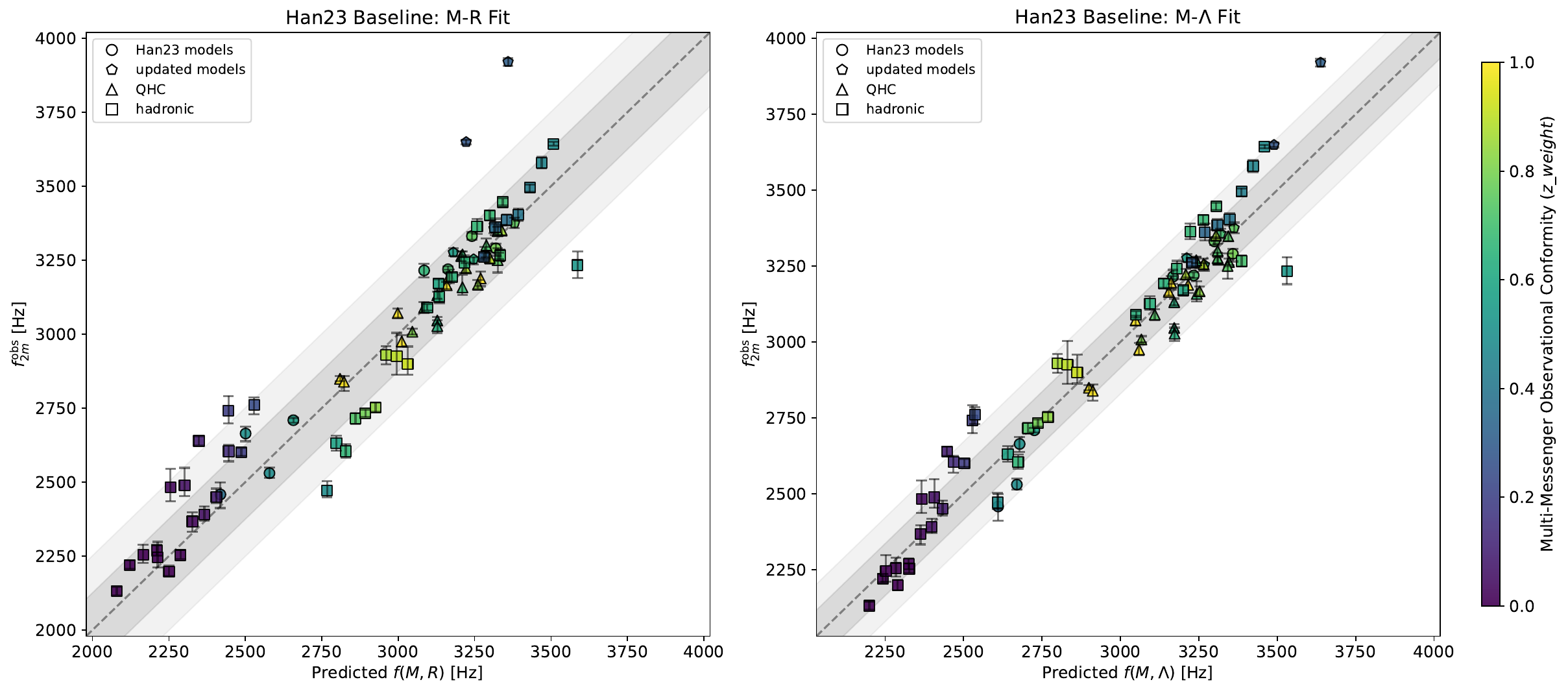}
\vspace{-0.3cm}
\caption{Simulated dominant post-merger frequency $f_{2,\rm mean}$ versus the value predicted by the quasi-universal relation anchored to the Han23 baseline ($R_{1.4}=12.42^{+0.78}_{-1.06}\,{\rm km}$, $\Lambda_{1.4}=465^{+224}_{-214}$), for the $M$--$R$ fit [left, Eq.~(\ref{eq:old_R})] and the $M$--$\ln\Lambda$ fit [right, Eq.~(\ref{eq:old_L})]. Marker shapes denote the EOS families, and the color encodes the multi-messenger conformity weight $w$ [Eq.~(\ref{eq:weight})], with warmer colors lying closer to the observationally favored region. The dashed line marks perfect agreement; the dark and light gray bands show the $\pm\sigma_{f_2}$ and $\pm2\sigma_{f_2}$ intervals, with $\sigma_{f_2}=125.6$ and $85.7\,{\rm Hz}$ for the two representations, respectively.}
\label{fig:fundamental_old}
\end{figure*}

\begin{figure*}[htbp]
\centering
\vspace{-0.3cm}
\includegraphics[width=1\textwidth]{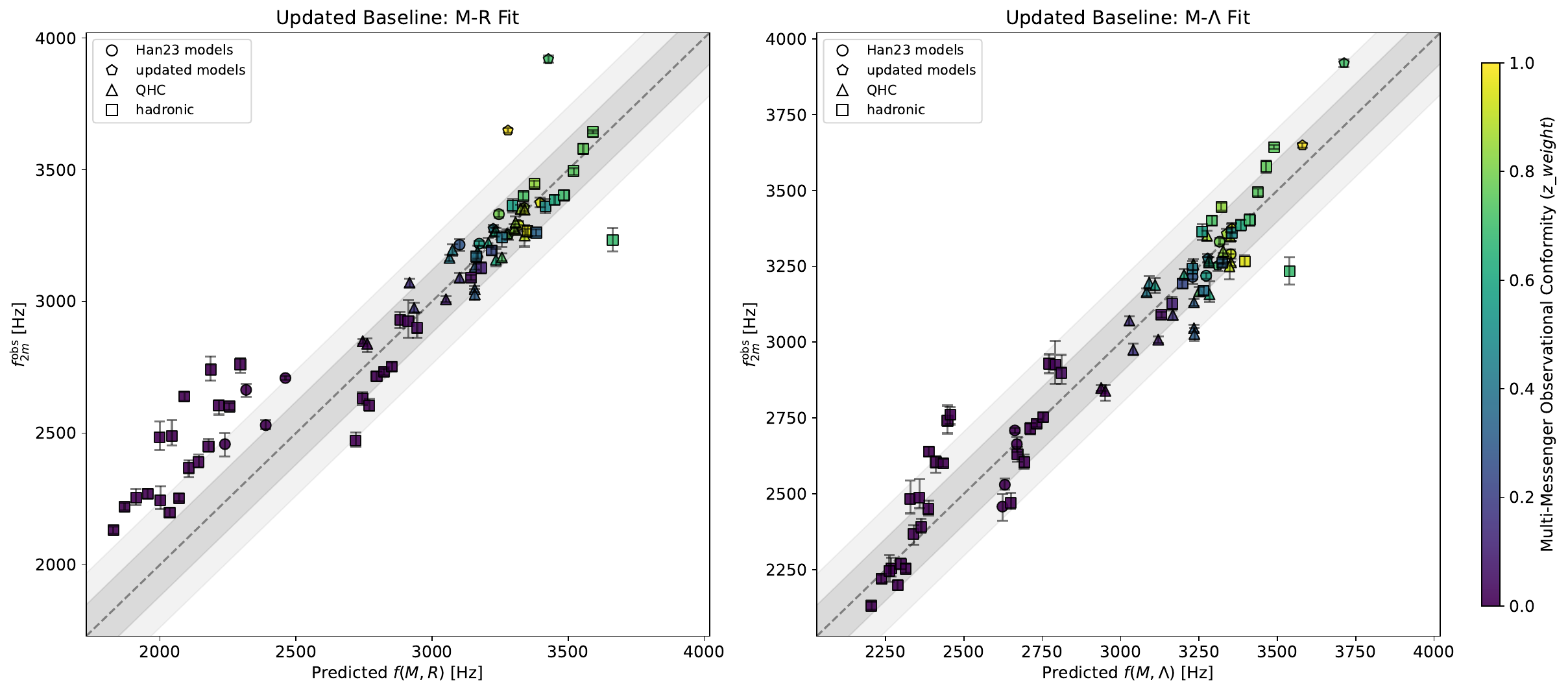}
\vspace{-0.3cm}
\caption{As in Fig.~\ref{fig:fundamental_old}, but anchored to the updated baseline ($R_{1.4}=11.55^{+0.50}_{-0.39}\,{\rm km}$, $\Lambda_{1.4}=272^{+96}_{-55}$), using Eqs.~(\ref{eq:new_R}) (left) and~(\ref{eq:new_L}) (right). The weighted dispersion about the relation is $\sigma_{f_2}=119.2\,{\rm Hz}$ ($M$--$R$) and $102.3\,{\rm Hz}$ ($M$--$\ln\Lambda$).}
\label{fig:fundamental_new}
\end{figure*}

Figures~\ref{fig:fundamental_old} and~\ref{fig:fundamental_new} compare the predicted and observed $f_{2,\rm mean}$ for the Han23 and updated baselines, respectively. Here, each baseline is anchored by the 68\% credible intervals of the $R$ and $\Lambda$ posteriors for a $1.4\,M_\odot$ neutron star. The color of each point encodes its observational conformity weight~$w$, and the gray bands mark the $\pm\sigma_{f_2}$ and $\pm 2\sigma_{f_2}$ intervals around the line of perfect agreement.

For the Han23 baseline ($R_{1.4}=12.42^{+0.78}_{-1.06}\,{\rm km}$, $\Lambda_{1.4}=465^{+224}_{-214}$), the best-fit relations are
\begin{equation}
    f_2^{\rm fit}(M,R) = 3013.9 + 1647.8\,(M-1.4) - 430.0\,(R-12.42),
    \label{eq:old_R}
\end{equation}
and
\begin{equation}
    f_2^{\rm fit}(M,\ln\Lambda) = 3054.3 - 2213.0\,(M-1.4) - 800.6\,\ln\frac{\Lambda}{465}.
    \label{eq:old_L}
\end{equation}

The radius-based fit yields a weighted RMSE of $\sigma_{f_2}=125.6\,{\rm Hz}$, whereas the $\Lambda$-based fit achieves a substantially smaller weighted RMSE of $\sigma_{f_2}=85.7\,{\rm Hz}$, indicating a tighter correlation.

For the updated baseline ($R_{1.4}=11.55^{+0.50}_{-0.39}\,{\rm km}$, $\Lambda_{1.4}=272^{+96}_{-55}$), the corresponding relations are
\begin{equation}
    f_2^{\rm fit}(M,R) = 3406.7 + 1555.3\,(M-1.4) - 556.2\,(R-11.55),
    \label{eq:new_R}
\end{equation}
and
\begin{equation}
    f_2^{\rm fit}(M,\ln\Lambda) = 3487.4 - 2982.0\,(M-1.4) - 871.0\,\ln\frac{\Lambda}{272}.
    \label{eq:new_L}
\end{equation}
Here, the $M$--$R$ fit gives a weighted RMSE of $\sigma_{f_2}=119.2\,{\rm Hz}$, while the $M$--$\Lambda$ fit results in a weighted RMSE of $\sigma_{f_2}=102.3\,{\rm Hz}$. The lower weighted RMSE of the $\Lambda$-based fit confirms that the tidal deformability captures the pre-merger compactness more effectively than the radius alone~\citep{Bauswein2012, Bernuzzi2015, Breschi2019} (the tidal deformability is also a direct observable in GW measurements, while statements on the radius require modelling through quasi-universal relations). We caution that $M$ and $\Lambda$ are themselves strongly anti-correlated across the ensemble, so that the individual coefficients $\alpha_M$ and $\alpha_\Lambda$ in the $M$--$\ln\Lambda$ representation are partially degenerate; it is their combination, rather than either coefficient in isolation, that carries a direct physical interpretation.

The $M$--$\Lambda$ fit achieves a lower $\sigma_{f_2}$ than the $M$--$R$ one in both baselines ($85.7 < 125.6\,{\rm Hz}$ and $102.3 < 119.2\,{\rm Hz}$), confirming that $\Lambda$, which integrates the EOS profile over the stellar interior rather than only sampling the surface, is a more efficient proxy for the second physical mechanism. The $M$--$\Lambda$ prediction uncertainty is lower for the Han23 baseline ($85.7\,{\rm Hz}$) than for the updated baseline ($102.3\,{\rm Hz}$), despite the new data providing tighter constraints overall. This fact reflects the influence of the third mechanism. The high-weight Han23 models occupy the stiff EOS regime ($\Lambda \gtrsim 400$, $R \gtrsim 12\,{\rm km}$), where the supranuclear EOS does not depart strongly from the behavior predicted by the sub-nuclear stiffness, and the $f_2$--$\Lambda$ relation is empirically tight. The high-weight updated models instead cluster in a softer regime ($\Lambda \approx 200$--$600$, $R \approx 11.2$--$12.1\,{\rm km}$) where the pQCD-driven cross-density trade-off of Fig.~\ref{fig:selected_models} causes the supranuclear stiffness to depart more significantly from the prediction based on $R$ or $\Lambda$ alone, producing the larger residual ($\sigma_{f_2} \approx 102\,{\rm Hz}$). 

The fact that we found that, after simultaneously controlling for binary mass and pre-merger compactness (either $\Lambda$ or $R$), the weighted $f_{2,\rm mean}$ uncertainty under the latest multi-messenger constraints is 
$\sigma_{f_2} \approx 100\,{\rm Hz}$ reflects the intrinsic dispersion of the cold EOS within the observationally constrained posterior. This indicates that the currently allowed range of supranuclear EOS behavior alone produces this level of variation. We note that this figure is a conservative upper bound on the pure cold-EOS dispersion, because $\sigma_{f_2}$ also absorbs the per-model frequency-extraction and finite-resolution uncertainties of the underlying simulations. Being drawn from independent codes and grid setups across the heterogeneous ensemble, these numerical errors enter as quasi-random fluctuations that average out of the fitted coefficients rather than biasing the central relation. Common-mode numerical shifts, in turn, largely cancel in the differential quantities (the $\Gamma_{\rm th}$ shift, the soft--stiff contrast, and the $(f_1+f_3)/2$ relation; see below) that underlie our conclusions, so a higher-resolution calculation would only tighten $\sigma_{f_2}$. 

An independent estimate of the thermal contribution is obtained by comparing the supplementary $\Gamma_{\rm th}=1.6$ simulations with the primary $\Gamma_{\rm th}=2.0$ results for the stiff updated models. This reveals a frequency shift of $\sim 100$--$120\,{\rm Hz}$ at $M_{\rm NS} = 1.30$--$1.35\,M_\odot$ (see also the similar results of \cite{Han:2025pho}). Since this thermal shift is comparable to $\sigma_{f_2}$, the primary fit based on $\Gamma_{\rm th}=2.0$ effectively sets a lower bound on $f_{2,\rm mean}$ at each $(M,\Lambda)$ or $(M,R)$. Consequently, a future post-merger measurement that lies noticeably above this bound would favor a lower value of $\Gamma_{\rm th}$, thereby providing evidence for stronger thermal softening and placing meaningful constraints on the finite-temperature behavior of dense matter.

\begin{figure*}[htbp]
\centering
\vspace{-0.3cm}
\includegraphics[width=1\textwidth]{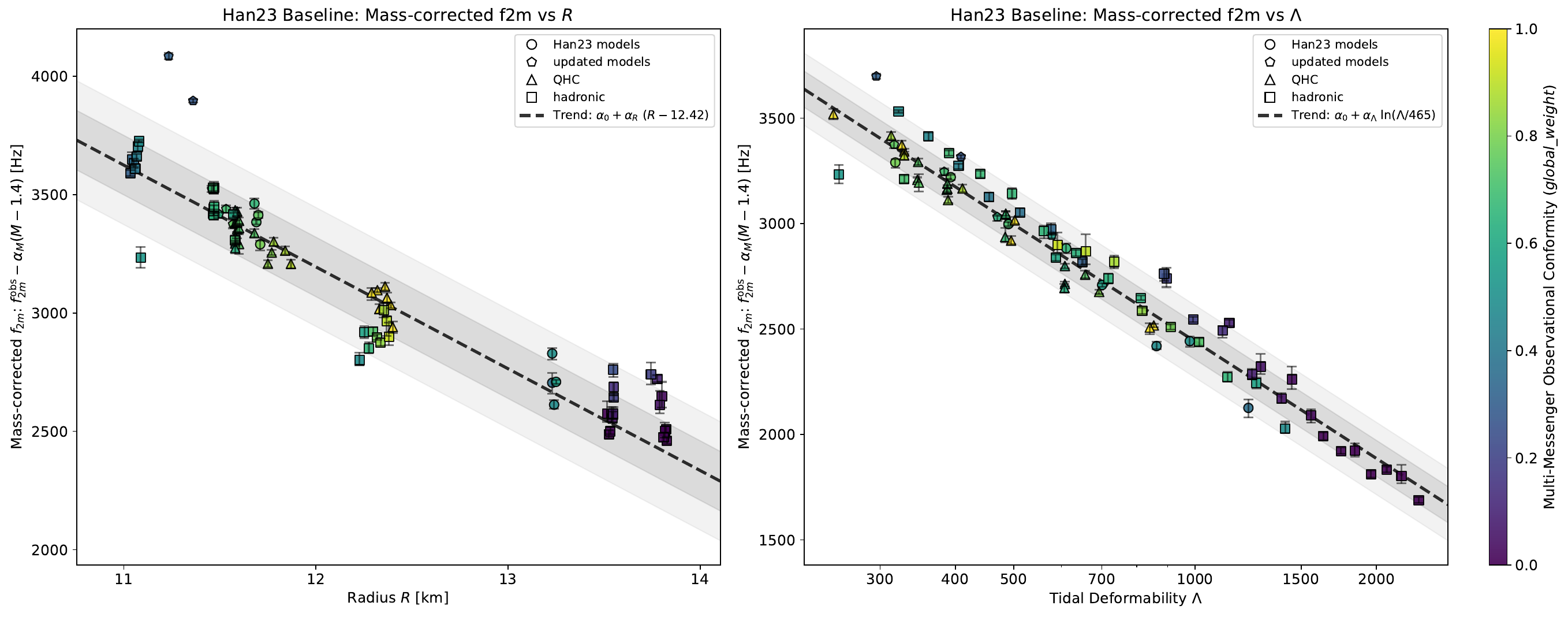}
\vspace{-0.3cm}
\caption{Mass-corrected dominant post-merger frequency $\delta f_2\equiv f_{2,\rm mean}-\alpha_M\,(M-1.4\,M_\odot)$ as a function of stellar radius $R$ (left) and tidal deformability $\Lambda$ (right) for the Han23 baseline. Removing the leading mass dependence isolates the influence of the pre-merger compactness and of the supranuclear EOS. Marker shapes, color scale, and gray bands are as in Fig.~\ref{fig:fundamental_old}; the dashed line is the best-fit trend.}
\label{fig:detrended_old}
\end{figure*}

\begin{figure*}[htbp]
\centering
\vspace{-0.3cm}
\includegraphics[width=1\textwidth]{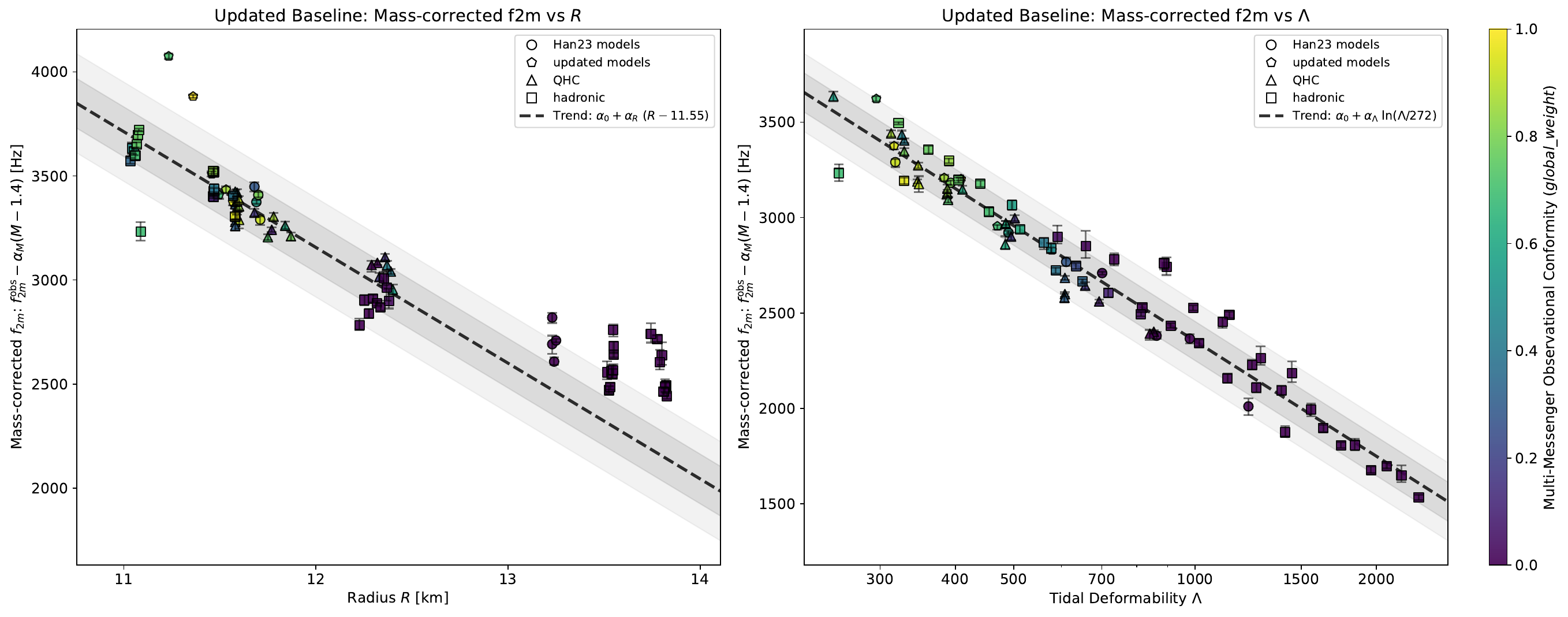}
\vspace{-0.3cm}
\caption{As in Fig.~\ref{fig:detrended_old}, but for the updated baseline. The high-weight (observationally favored) models cluster tightly within $\pm\sigma_{f_2}$ of the trend, whereas the low-weight models, corresponding to very stiff ($R\gtrsim12.5\,{\rm km}$) or very soft ($R\lesssim11\,{\rm km}$) EOSs now disfavored by multi-messenger data, deviate by more than $500\,{\rm Hz}$, visualizing the reduction of the plausible $f_{2,\rm mean}$ range from $\gtrsim500\,{\rm Hz}$ to $\sim100\,{\rm Hz}$.}
\label{fig:detrended_new}
\end{figure*}

By defining the mass-corrected frequency $\delta f_2 \equiv f_{2,\rm mean} - \alpha_M\,(M-1.4\,M_\odot)$, we remove the leading mass dependence and isolate the influence of compactness and supranuclear physics in the $\delta f_2$--$R$ and $\delta f_2$--$\ln\Lambda$ planes (Figs.~\ref{fig:detrended_old} and~\ref{fig:detrended_new}).

In the updated-baseline mass-corrected plane (Fig.~\ref{fig:detrended_new}), the high-weight models cluster tightly within $\pm\sigma_{f_2}$ of the best-fit trend, whereas low-weight models, which are very stiff ($R > 12.5\,{\rm km}$) or very soft ($R < 11\,{\rm km}$) and now disfavored by current data, deviate by more than $500\,{\rm Hz}$. This contrast confirms that multi-messenger observations have reduced the plausible $f_{2,\rm mean}$ uncertainty from $\gtrsim 500\,{\rm Hz}$ to $\sim 100\,{\rm Hz}$. For the stiff updated EOS, comparing the supplementary $\Gamma_{\rm th}=1.6$ runs to the primary $\Gamma_{\rm th}=2.0$ results using the values listed in Table~\ref{tab:allmodels} reveals that the $\Gamma_{\rm th}=1.6$ models lie systematically above their $\Gamma_{\rm th}=2.0$ counterparts by $\sim 100$--$120\,{\rm Hz}$ at $M_{\rm NS} = 1.30$--$1.35\,M_\odot$. Since $\Gamma_{\rm th}$ parametrizes only the thermal part of the EOS while leaving the cold part unchanged, this shift directly measures the finite-temperature contribution at the currently-allowed EOS stiffness. The fact that this thermal shift is of the same order as the cold-EOS dispersion $\sigma_{f_2}$ indicates that these two sources of uncertainty are comparable, and both must be accounted for when interpreting a future post-merger frequency measurement. If a significantly larger high-frequency deviation were observed, it could point to a finite-temperature phase transition, because a transition point that shifts to lower densities at finite temperature would produce an even more compact remnant and a correspondingly higher $f_{2,\rm mean}$.

A future measurement of $f_{2,\rm mean}$ to $\lesssim 100\,{\rm Hz}$ precision, which is within the projected sensitivity of Einstein Telescope~\cite{Punturo2010} or Cosmic Explorer~\cite{Reitze2019} for nearby events~\citep{Prakash2024PhRvD109}, combined with an inspiral determination of $(M, \Lambda)$, would therefore yield a direct constraint on the thermal EOS of supranuclear matter, a regime entirely inaccessible to stable neutron-star observations. In this sense, the $\sim 100\,{\rm Hz}$ $\sigma_{f_2}$ is not a limitation to be minimised but a scientifically informative quantity, representing the unique contribution that post-merger GW observations would add to our knowledge of dense matter at finite temperature.

Although the precise functional forms of the nonlinear couplings that generate the GW secondary peaks remain to be established, the observed symmetry of the measured peaks implies that $f_1$ and $f_3$ depend on $f_2$ and $f_0$ in such a way that $(f_1+f_3)/2 \approx f_{2,\rm mean}$ holds to good accuracy~\citep{Sterg-2011MNRAS.418, Takami-Rezzolla-Baiotti-2014PhRvL, Takami-Rezzolla-Baiotti-PhRvD91, Bauswein-Stergioulas-2015PhRvD91}, a relation that can be tested directly against the post-merger GW spectrum.

\begin{figure}[htbp]
\centering
\vspace{-0.3cm}
\includegraphics[width=0.5\textwidth]{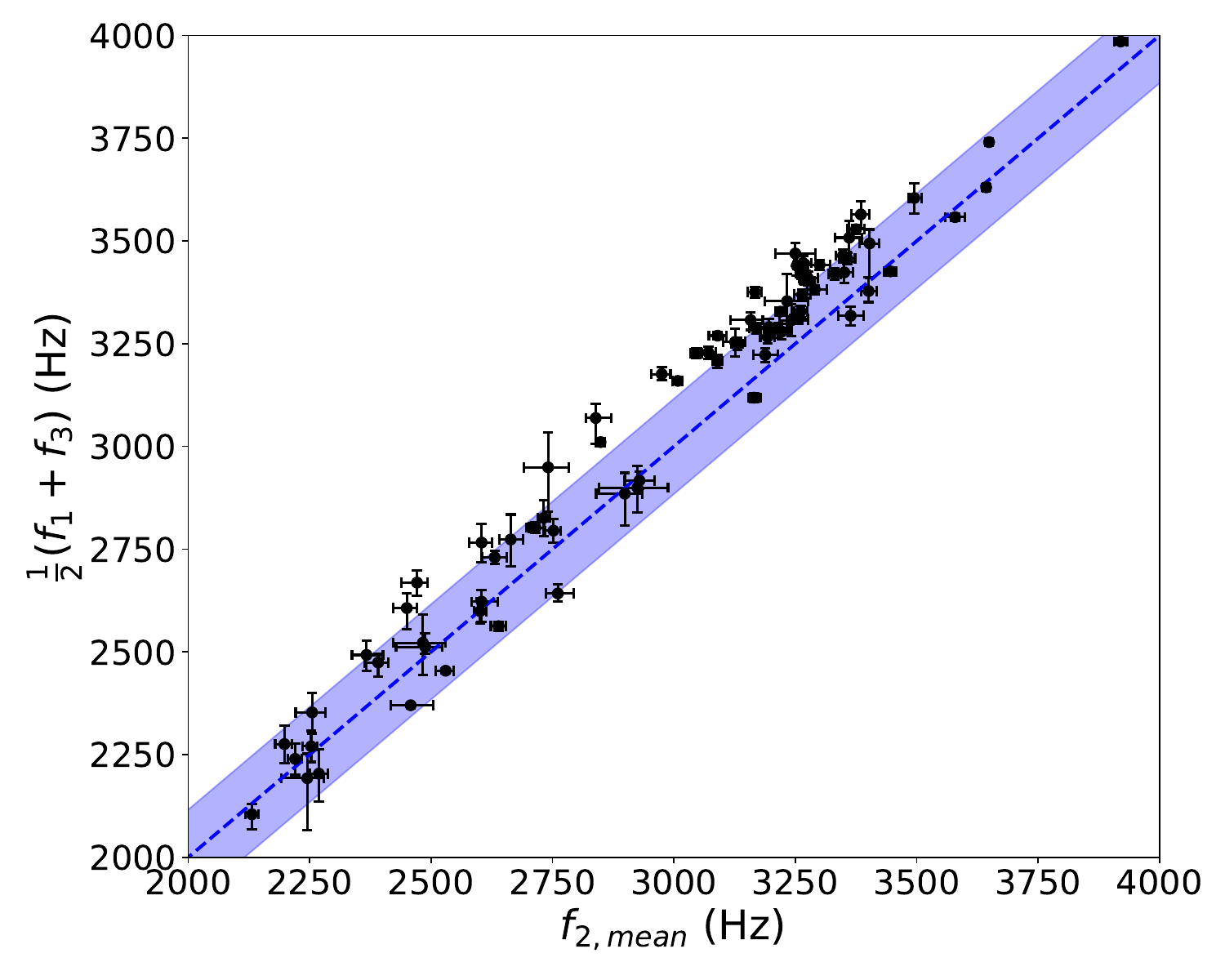}
\vspace{-0.3cm}
\caption{Half-sum of the secondary post-merger peaks, $(f_1+f_3)/2$, versus the dominant peak $f_{2,\rm mean}$ for all primary-sample models in which $f_1$ and $f_3$ can be unambiguously identified. The dashed line is the identity relation $(f_1+f_3)/2=f_{2,\rm mean}$ expected from the quasi-linear coupling $f_{1,3}\approx f_{2,\rm mean}\mp f_0$, and the blue band marks its root-mean-square deviation $\sigma_{\rm couple}=116.0\,{\rm Hz}$. Error bars denote the 68\% confidence intervals of the extracted frequencies.}
\label{fig:f2withf1f3}
\end{figure}

Figure~\ref{fig:f2withf1f3} tests this relation for all primary-sample models in which $f_1$ and $f_3$ peaks can be clearly identified in the simulated GW spectrum. The relation holds across all EOS families, masses $M_{\rm NS} = 1.20$--$1.55\,M_\odot$, and $\Gamma_{\rm th}$ values with a RMSE of $\sigma_{\rm couple} = 116.0\,{\rm Hz}$, where $\sigma_{\rm couple}$ denotes the error of $(f_1+f_3)/2$ about $f_{2,\rm mean}$. This $\sigma_{\rm couple}$ is comparable to the $\sim 100$--$120\,{\rm Hz}$ $\sigma_{f_2}$ from the $f_2(M,R)$ or $f_2(M,\Lambda)$ fit, and both are set by the same physical scale. The consistency between these two independent estimates of the $\sim 100\,{\rm Hz}$ scale provides a cross-check that this scale characterises the post-merger physics uncertainty under current EOS constraints. 

From a detection standpoint, thanks to this universality, a joint measurement of $f_1$, $f_2$, and $f_3$ would enable a consistency check, because any departure $|(f_1+f_3)/2 - f_{2,\rm mean}| \gg 116\,{\rm Hz}$ would signal anomalous remnant dynamics such as strong magnetic braking or efficient angular-momentum redistribution. 

\section{Summary and Conclusions}
\label{sec:summary}

We have quantified the uncertainty in the dominant post-merger GW frequency $f_{2,\rm mean}$ of binary neutron star mergers that survives once the cold dense-matter EOS is constrained by the latest multi-messenger data. Instead of a small set of phenomenological EOSs, we sampled non-parametric EOS posteriors conditioned on GW tidal measurements, NICER mass--radius data, massive-pulsar masses, and perturbative-QCD constraints at asymptotically high density. For each binary mass we evolved the softest and stiffest members of the multi-messenger posterior with fully general-relativistic hydrodynamic simulations, bracketing the thermal response through an ideal-gas component with $\Gamma_{\rm th}=1.6$ and $2.0$, and we analyzed them as part of a larger sample of $82$ models spanning hadronic, quark--hadron crossover, and statistical EOS families.

We obtain three main results. First, the evolution of the maximum rest-mass density shows that the survival of the remnant is governed jointly by the pre-merger compactness and the thermal prescription, with the compact soft branch of the pQCD-constrained ensemble collapsing at progressively lower masses as $\Gamma_{\rm th}$ decreases, so that for $M_{\rm tot}\gtrsim2.70\,M_\odot$ a detectable $f_2$ signal already selects the stiffer high-density branch. Second, after simultaneously controlling for binary mass and a pre-merger compactness proxy ($\Lambda$ or $R$) through a Gaussian-weighted quasi-universal fit, the residual prediction uncertainty is $\sigma_{f_2}\simeq100\,{\rm Hz}$, a factor of several below the $\gtrsim500\,{\rm Hz}$ spread found considering also EOSs no longer favored by current data. The tidal deformability is a more efficient compactness proxy than the radius, and the larger residual obtained for the updated baseline reflects the pQCD-enforced anti-correlation between low- and high-density stiffness, which partially decouples the supranuclear EOS from inspiral observables. Third, the secondary spectral peaks obey $(f_1+f_3)/2\approx f_{2,\rm mean}$ to $\sigma_{\rm couple}=116\,{\rm Hz}$ across all EOS families, masses, and thermal indices, providing both a consistency check and a model-independent estimator of $f_{2,\rm mean}$.

The residual cold-EOS uncertainty of $\sim100\,{\rm Hz}$ is comparable to the $\sim100$--$120\,{\rm Hz}$ frequency shift induced by varying $\Gamma_{\rm th}$ from $2.0$ to $1.6$. Because the $\Gamma_{\rm th}=2.0$ fit effectively sets a lower bound on $f_{2,\rm mean}$ at fixed $(M,\Lambda)$ or $(M,R)$, future measurements lying systematically above this bound would favor stronger thermal softening~\cite{Blacker-2024PhRvD.109,Kochankovski-2025PhRvD.112} or, if the excess were large, a finite-temperature phase transition shifting to lower density at high temperature~\citep{Prakash2024PhRvD109}. A post-merger detection at $\lesssim100\,{\rm Hz}$ precision, within the projected reach of the Einstein Telescope and Cosmic Explorer for nearby events, combined with an inspiral determination of $(M,\Lambda)$, would therefore directly constrain the finite-temperature EOS of supranuclear matter, a regime inaccessible to observations of cold, stable neutron stars.

Taken together, these results establish the dominant post-merger frequency as a quantitative, multi-messenger-calibrated probe of supranuclear matter that complements inspiral and pulsar observations of cold neutron stars. In the era of next-generation detectors, applying this framework to the growing catalog of kilohertz post-merger signals will open a direct observational window on the finite-temperature equation of state and on a possible hadron--quark transition in the remnant core.

\acknowledgments
We thank Kentaro Takami for providing the GW waveforms for hadronic models, and Ming-Zhe Han for providing several equations of state used in our simulations. Simulations were performed on the Hokusai Bigwaterfall supercomputer in RIKEN. Y.H is supported by the Postdoctoral Fellowship Program (No. GZC20241915) of the China Postdoctoral Science Foundation, the Project for Young Scientists in Basic Research (No. YSBR-088) of the Chinese Academy of Sciences, and the National Natural Science Foundation of China under Grants (No. 12588101 and No. 12233011). L.B. is supported by the JSPS KAKENHI grant No. JP25H00675.

\clearpage

\bibliographystyle{apsrev4-2}
\bibliography{ref.bib}{}

\end{document}